\documentclass{article}
\usepackage{amssymb}
\usepackage{amsfonts}
\usepackage{graphicx}

\input{tcilatex}

\begin{document}

\title{Non-differentiable Bohmian trajectories}
\author{Gebhard Gr\"{u}bl and Markus Penz \\
Theoretical Physics Institute, Universit\"{a}t Innsbruck,\\
Technikerstr 25, A-6020 Innsbruck Austria}
\maketitle

\begin{abstract}
A solution $\psi $ to Schr\"{o}dinger's equation needs some degree of
regularity in order to allow the construction of a Bohmian mechanics from
the integral curves of the velocity field $\hbar \Im \left( \bigtriangledown
\psi /m\psi \right) .$ In the case of one specific non-differentiable weak
solution $\Psi $ we show how Bohmian trajectories can be obtained for $\Psi $
from the trajectories of a sequence $\Psi _{n}\rightarrow \Psi .$ (For any
real $t$ the sequence $\Psi _{n}\left( t,\cdot \right) $ converges
strongly.) The limiting trajectories no longer need to be differentiable.
This suggests a way how Bohmian mechanics might work for arbitrary initial
vectors $\Psi $ in the Hilbert space on which the Schr\"{o}dinger evolution $%
\Psi \mapsto e^{-iht}\Psi $ acts.
\end{abstract}

\section{Introduction}

Quantum mechanis often is praised as a theory which unifies classical
mechanics and classical wave theory. Quanta are said to behave either as
particles or waves, depending on the type of experiment they are subjected.
But where in the standard formalism can the particles of the interpretive
talk be found? Perhaps only to some degree in the reduction postulate
applied to position measurements. In reaction to this unsatisfactory state
of affairs, Bohmian mechanics introduces a mathematically precise particle
concept into quantum mechanical theory. The fuzzy wave functions are
supplemented by sharp particle world lines. Through this additional
structure some quantum phenoma like the double slit experiment have lost
their mystery.

Clearly the additional structure of particle world lines brings along its
own mathematical problems. Ordinary differential equations are generated
from solutions of partial differential equations. A mathematically
convincing general treatment so far has been given for a certain type of
wave functions which do not exhaust all possible quantum mechanical
situations. Exactly this fact has led some workers to doubt that a Bohmian
mechanics exists for all initial states $\Psi _{0}$ of a Schr\"{o}dinger
evolution $t\mapsto e^{-iht}\Psi _{0}.$ We shall show on one specific case
of a counter example $\Psi _{0}$ how the problem might be resolved in
general. We approximate the state $\Psi _{0},$ for which the Bohmian
velocity field does not exist, by states which do have one. Their integral
curves turn out to converge to limit curves which can be taken to constitute
the Bohmian mechanics of the state unamenable to Bohmian mechanics on first
sight.

\section{Bohmian evolution for $\protect\psi \in C^{2}$}

Let $\psi :\mathbb{R}\times \mathbb{R}^{s}\rightarrow \mathbb{C}$ be twice
continuously differentiable, i.e. $\psi \in C^{2}\left( \mathbb{R}\times
\mathbb{R}^{s}\right) ,$ and let $\psi $ obey Schr\"{o}dinger's partial
differential equation%
\begin{equation}
i\hbar \partial _{t}\psi \left( t,x\right) =-\frac{\hbar ^{2}}{2m}\Delta
\psi \left( t,x\right) +V\left( x\right) \psi \left( t,x\right)
\label{Schroed}
\end{equation}%
with $V:\mathbb{R}^{s}\rightarrow \mathbb{R}$ being smooth, i.e., $V\in
C^{\infty }\left( \mathbb{R}^{s}\right) .$ From $\psi ,$ which is called a
classical solution of Schr\"{o}dinger's equation, a deterministic time
evolution $x\mapsto \gamma _{x}\left( t\right) $ of certain points $x\in
\mathbb{R}^{s}$ can be derived: If there exists a unique maximal solution $%
\gamma _{x}:I_{x}\rightarrow \mathbb{R}^{s}$ to the implicit first order
system of ordinary differential equations%
\begin{equation}
\rho _{\psi }\left( t,\gamma \left( t\right) \right) \dot{\gamma}\left(
t\right) =j_{\psi }\left( t,\gamma \left( t\right) \right)  \label{BMmotion}
\end{equation}%
with the initial condition $\gamma \left( 0\right) =x,$ one takes $\gamma
_{x}$ as the evolution of $x.$ Here $\rho _{\psi }:\mathbb{R}\times \mathbb{R%
}^{s}\rightarrow \mathbb{R}$ and $j_{\psi }:\mathbb{R}\times \mathbb{R}%
^{s}\rightarrow \mathbb{R}^{s}$ with%
\begin{equation}
\rho _{\psi }\left( t,x\right) =\left\vert \psi \left( t,x\right)
\right\vert ^{2}\text{ and }j_{\psi }\left( t,x\right) =\frac{\hbar }{m}\Im %
\left[ \overline{\psi \left( t,x\right) }\nabla _{x}\psi \left( t,x\right) %
\right]  \label{Current}
\end{equation}%
obey the continuity equation $\partial _{t}\rho _{\psi }\left( t,x\right) =-$%
div$j_{\psi }\left( t,x\right) $ for all $\left( t,x\right) \in \mathbb{R}%
\times \mathbb{R}^{s}.$ From now on we shall drop the index $\psi $ from $%
\rho _{\psi }$ and $j_{\psi }.$

For certain solutions\footnote{%
The simplest explicitly solvable example is provided by the plane wave
solution $\psi \left( t,x\right) =e^{-i\left\vert k\right\vert ^{2}t+ik\cdot
x}.$ Its Bohmian evolution $\Phi $ obeys $\Phi \left( t,x\right) =tk.$
Another explicitly solveable case is given by a Gaussian free wave packet.} $%
\psi $ the curves $\gamma _{x}$ can be shown to exist on a maximal domain $%
I_{x}=\mathbb{R}$ for all $x\in \mathbb{R}^{s}:$ \emph{If} $\psi $ has no
zeros, then the velocity field $v=j/\rho $ is a $C^{1}$-vector field. $v$
then obeys a local Lipschitz condition such that the maximal solutions are
unique. \emph{If} in addition there exist continuous nonnegative real
functions $\alpha ,\beta $ with $\left\vert v\left( t,x\right) \right\vert
\leq \alpha \left( t\right) \left\vert x\right\vert +\beta \left( t\right) $
then all maximal solutions equation (\ref{BMmotion}) are defined on $\mathbb{%
R}$ and the general solution
\[
\Phi :\bigcup\nolimits_{x\in \mathbb{R}^{s}}I_{x}\times \left\{ x\right\}
\rightarrow \mathbb{R}^{s}\text{ with }\Phi \left( t,x\right) =\gamma
_{x}\left( t\right)
\]%
extends to all of $\mathbb{R}\times \mathbb{R}^{s}.$ (Thm 2.5.6, ref. \cite%
{Aul}) Due to the uniqueness of maximal solutions the map $\Phi \left(
t,\cdot \right) :\mathbb{R}^{s}\rightarrow \mathbb{R}^{s}$ is a bijection
for all $t\in \mathbb{R}.$ It obeys\footnote{%
Here $\Phi \left( t,\Omega \right) =\left\{ \Phi \left( t,x\right)
\left\vert x\in \Omega \right. \right\} .$}%
\begin{equation}
\int_{\Phi \left( t,\Omega \right) }\rho \left( t,x\right)
d^{s}x=\int_{\Omega }\rho \left( 0,x\right) d^{s}x  \label{Tranport}
\end{equation}%
for all $t\in \mathbb{R}$ and for all open subsets $\Omega \subset \mathbb{R}%
^{s}$ with sufficiently smooth boundary such that the integral theorem of
Gauss can be applied to the space time vector field $\left( \rho ,j\right) $
on the domain $\bigcup\nolimits_{t^{\prime }\in \left( 0,t\right) }\Phi
\left( t^{\prime },\Omega \right) .$ \cite{DTe}

These undisputed mathematical facts have instigated Bohm's amendment of
equation (\ref{Schroed}) in order to explain the fact that \emph{macroscopic
bodies usually are localized much stricter than their wave functions suggest.%
}

In Bohm's completion of nonrelativistic quantum mechanics it is assumed that
any closed system has at any time, in addition to its wave function, a
position in its configuration space and that this position evolves according
to the general solution $\Phi $ induced by the wave function. One says that
the position is guided by $\psi $ since $\Phi $ is completely determined by $%
\psi $ (and no other forces than the ones induced by $\psi $ are allowed to
act on the position).$\ $More specifically, $\gamma _{x}$ is assumed to give
the position evolution for an isolated particle with wave function $\psi
\left( 0,\cdot \right) $ and position $x$ -- both at time $t=0.$

As is common in standard quatum mechanics, $\psi \left( 0,\cdot \right) $ is
supposed to obey%
\[
\int_{\mathbb{R}^{s}}\left\vert \psi \left( 0,x\right) \right\vert
^{2}d^{s}x=1.
\]%
The nonnegative density $\rho \left( 0,\cdot \right) $ is interpreted as the
probability density of the position which the particle has at time $t=0.$
Since an initial position $x$ is assumed to evolve into $\gamma _{x}\left(
t\right) ,$ the position probability density at time $t$ is then, due to
equation (\ref{Tranport}), given by $\rho \left( t,\cdot \right) .$ In
particular, Bohm's completion gives the position probabilities among all the
other spectral probability measures a fundamental status, since the
empirical meaning of the other ones, as for instance momentum probabilities,
all are deduced from position probabilities.

There are classical solutions of Schr\"{o}dinger's equation, whose general
solution $\Phi $ \emph{does not extend} to all of $\mathbb{R}\times \mathbb{R%
}^{s}.$ An obstruction to do so can be posed by the zeros of $\psi .$ In the
neigbourhood of such zero the velocity field $v=j/\rho $ may be unbounded
and $v$ then lacks a continuous extension into the zero. As an example
consider a time $0$ wave function $\psi \left( 0,\cdot \right) :\mathbb{R}%
^{2}\rightarrow \mathbb{C},$ for which $\psi \left( 0,x,y\right)
=x^{2}+iy^{2}$ within a neighbourhood $U$ of its zero $\left( x,y\right)
=\left( 0,0\right) .$ Within $U$ for the velocity field follows
\[
\frac{m}{\hbar }v^{1}\left( 0,x,y\right) =\Im \frac{\partial _{x}\psi \left(
0,x,y\right) }{\psi \left( 0,x,y\right) }=-\frac{2xy^{2}}{x^{4}+y^{4}}.
\]%
Hence for $0<\left\vert \phi \right\vert <\pi /2$ we have $v^{1}\left( r\cos
\phi ,r\sin \phi \right) \rightarrow -\infty $ for $r\rightarrow 0$ with $%
\phi $ fixed. Thus the implicit Bohmian evolution equation (\ref{BMmotion})
is singular in a zero of the wave function whenever the velocity field does
not have a continuous extension into it. As a consequence the evolution $%
\gamma _{x}$ of such a zero $x$ is not defined by equation (\ref{BMmotion}).

As a related phenomenon there are solutions to equation (\ref{BMmotion})
which begin or end at a finite time because they terminate at a zero of $%
\psi .$ A nice example \cite{Bndl} for this to happen provide the zeros of
the harmonic oscillator wave function $\psi :\mathbb{R}^{2}\rightarrow
\mathbb{C}$ with%
\[
\psi \left( t,x\right) =e^{-\frac{x^{2}}{2}}\left( 1+e^{-2it}\left(
1-2x^{2}\right) \right) .
\]%
E.g., the points $\left\vert x\right\vert =1$ are zeros of $\psi \left(
t,\cdot \right) $ at the times $t\in \pi \mathbb{Z}.$ They are singularities
of $v$ since%
\[
\lim_{t\rightarrow 0}t\Im \frac{\partial _{x}\psi \left( t,\pm 1\right) }{%
\psi \left( t,\pm 1\right) }=\pm 2.
\]%
Note however that $\Im \frac{\partial _{x}\psi \left( 0,x\right) }{\psi
\left( 0,x\right) }=0$ for $x\neq \pm 1.$

There are more challenges to Bohmian mechanics. The notion of distributional
solutions to a partial differential equation like (\ref{Schroed}) raises the
question whether these solutions support a kind of Bohmian particle motion
like the classical solutions do. After all quantum mechanics employs such
distributional solutions.

\section{Bohmian evolution for $\Psi _{t}\in C_{h}^{\infty }$}

In standard quantum mechanics the classical solutions, i.e. the $C^{2}$%
-solutions of equation (\ref{Schroed}), do not represent all physically
possible situations. Rather a more general quantum mechanical evolution is
abstracted from equation (\ref{Schroed}). It is given by the socalled weak
solutions%
\[
\Psi _{0}\mapsto \Psi _{t}=e^{-iht}\Psi _{0}\text{ for all }\Psi _{0}\in
L^{2}\left( \mathbb{R}^{s}\right)
\]%
with $\hbar h$ being a self-adjoint, ususally unbounded hamiltonian
corresponding to equation (\ref{Schroed}). The domain $D_{h}$ of $h$ does
not comprise all of $L^{2}\left( \mathbb{R}^{s}\right) ,$ yet it is dense in
$L^{2}\left( \mathbb{R}^{s}\right) .$ Since $h$ is self-adjoint, the
exponential $e^{-iht}$ has a unique continuous extension to $L^{2}\left(
\mathbb{R}^{s}\right) .$ This unitary evolution operator $e^{-iht}$
stabilizes the domain of $h$ as a dense subspace of $L^{2}\left( \mathbb{R}%
^{s}\right) .$ Thus if and only if an initial vector $\Psi _{0}$ belongs to $%
D_{h},$ equation (\ref{Schroed}) generalizes to
\begin{equation}
\lim_{\varepsilon \rightarrow 0}\left\Vert i\frac{\Psi _{t+\varepsilon
}-\Psi _{t}}{\varepsilon }-h\Psi _{t}\right\Vert =0  \label{SchroedQM}
\end{equation}%
for all $t\in \mathbb{R}.$ For $\Psi _{0}\notin D_{h}$ equation (\ref%
{SchroedQM}) does not hold for any time.

Yet the construction of Bohmian trajectories needs much more than the
evolution $\Psi _{0}\mapsto \Psi _{t}$ within $L^{2}\left( \mathbb{R}%
^{s}\right) ,$ since the elements of $L^{2}\left( \mathbb{R}^{s}\right) $
are equivalence classes $\left[ f\right] $ of functions $f\in \mathcal{L}%
^{2}\left( \mathbb{R}^{s}\right) .$ It rather needs a trajectory of
functions instead of a trajectory of equivalence classes of
square-integrable functions. \emph{If} there exists a function $\psi \in
C^{1}\left( \mathbb{R}\times \mathbb{R}^{s}\right) $ such that $\Psi \left(
t,\cdot \right) =\left[ \psi \left( t,\cdot \right) \right] $ holds for all $%
t\in \mathbb{R},$ then $\psi $ is unique and the Bohmian equation of motion (%
\ref{BMmotion}) can be derived from the evolution $\Psi _{0}\mapsto \Psi
_{t} $ through $\psi .$ When does there exist such $\psi ?$

Due to Kato's theorem (e.g. Thm X.15 of ref. \cite{RS}) the Schr\"{o}dinger
hamiltonians $h,$ corresponding to potentials $V$ from a much wider class
than just $C^{\infty }\left( \mathbb{R}^{s}:\mathbb{R}\right) ,$ have the
same domain as the free hamiltonian $-\Delta ,$ namely the Sobolev space $%
W^{2}\left( \mathbb{R}^{s}\right) .$ This is the space of all those $\Psi
\in L^{2}\left( \mathbb{R}^{s}\right) $ which have all of their
distributional partial derivatives up to second order being regular
distributions belonging to $L^{2}\left( \mathbb{R}^{s}\right) .$ Since $%
D_{h} $ is stabilized by the evolution $e^{-iht},$ for any $\Psi _{0}\in
D_{h}$ there exists for any $t\in \mathbb{R}$ a function $\psi \left(
t,\cdot \right) \in \mathcal{W}^{2}\left( \mathbb{R}^{s}\right) $ such that%
\begin{equation}
e^{-iht}\Psi _{0}=\left[ \psi \left( t,\cdot \right) \right] .  \label{Repr}
\end{equation}%
However, for this family of time parametrized functions $\psi \left( t,\cdot
\right) $ the Bohmian equation of motion in general does not make sense
since $\psi \left( t,\cdot \right) $ need not be differentiable in the
classical sense.\footnote{%
Only for $s=1$ Sobolev's lemma (Thm IX.24 in Vol 2 of ref. \cite{RS}) says
that $\left[ \psi \left( t,\cdot \right) \right] $ has a $C^{1}$
representative within $\mathcal{L}^{2}\left( \mathbb{R}\right) .$ From such
a $C^{1}$ representative $\psi \left( t,\cdot \right) $ the current $j$
follows as a continuous vector field and a continuous velocity field $v$ can
be derived outside the zeros of $\psi .$ However, $v$ does not need to obey
the local Lipschitz condition implying the local uniquenss of its integral
curves.}

Therefore some stronger restriction of initial data than $\Psi _{0}\in D_{h}$
is needed in order to supply the state evolution $\Psi _{0}\mapsto
e^{-iht}\Psi _{0}$ with Bohm's amendment. For a restricted set of initial
states $\left( x,\Psi _{0}\right) $ and for a fairly large class of static
potentials a$\ $Bohmian evolution has indeed been constructed in \cite{BDG}
and \cite{TeTu}. There it is shown that for any $\Psi _{0}\in
\bigcap\nolimits_{n\in \mathbb{N}}D_{h^{n}}=:C_{h}^{\infty }$ there exists

\begin{itemize}
\item a (time$\ $independent) subset $\Omega \subset \mathbb{R}^{s}$

\item for any $t$ a square-integrable function $\psi \left( t,\cdot \right) $
\end{itemize}

such that the restriction of $\psi \left( t,\cdot \right) $ to $\Omega $
belongs to $C^{\infty }\left( \Omega \right) $ and equation (\ref{Repr})
holds. The set $\Omega $ is obtained by removing from $\mathbb{R}^{s}$ first
the points where the potential function $V$ is not $C^{\infty },$ second the
zeros of $\psi \left( 0,\cdot \right) ,$ and third those points $x$ for
which the maximal solution $\gamma _{x}$ does not have all of $\mathbb{R}$
as its domain. Surprizingly, $\Omega $ is still sufficiently large, since
\[
\int_{\Omega }\left\vert \psi \left( 0,x\right) \right\vert ^{2}d^{s}x=1.
\]%
On this reduced set $\Omega $ of initial conditions a Bohmian evolution $%
\Phi :\mathbb{R}\times \Omega \rightarrow \mathbb{R}^{s}$ can be
constructed. Thus if $\Psi _{0}\in C_{h}^{\infty }$ and if the initial
position $x$ is distributed within $\mathbb{R}^{s}$ with probability density
$\left\vert \psi \left( 0,\cdot \right) \right\vert ^{2}$ then the global
Bohmian evolution $\gamma _{x}$ of $x$ exists with probability $1.$

\section{Bohmian evolution for $\Psi _{t}\in L^{2}\smallsetminus
C_{h}^{\infty }$}

How about initial conditions $\Psi _{0}\in L^{2}\left( \mathbb{R}^{s}\right)
\smallsetminus C_{h}^{\infty }?$ Can the equation of motion (\ref{BMmotion})
still be associated with $\Psi _{0}?$ Hall has devised a specific
counterexample $\Psi _{0}\notin D_{h}$ which leads to a wave function $\psi $
which at certain times is nowhere differentiable with respect to $x$ and
thus renders impossible the formation of the velocity field $v.$ Therefore
it has been brought forward that the Bohmian amendment of standard quantum
mechanics is \textquotedblleft formally incomplete\textquotedblright\ and it
has been claimed that the problem is unlikely to be resolved. \cite{Hal}

A promising way to tackle the problem is to succesively approximate the
initial condition $\Psi _{0}\notin D_{h}$ by a strongly convergent sequence
of vectors $\left( \Psi _{0}^{n}\right) \in C_{h}^{\infty }.$ For each of
the vectors $\Psi _{0}^{n}$ a Bohmian evolution $\Phi _{n}$ exists. We do
not know whether it has actually been either disproven or proven that the
sequence of evolutions does converge to a limit $\Phi $ and that the limit
depends on the chosen sequence $\Psi _{0}^{n}\rightarrow \Psi _{0}.$

Here we shall explore this question within the simplified setting of a
spatially one dimensional example. We will make use of an equation for $%
\gamma _{x}$ which has already been pointed out in \cite{BDG} and which does
not rely on the differentiability of $j.$ In this case equation (\ref%
{Tranport}) can be generalized in order to determine a nondifferentiable
Bohmian trajectory $\gamma _{x}$ by choosing $\Omega =\left( -\infty
,x\right) $ in (\ref{Tranport}) as follows.

Consider first the case of a $C^{2}$-solution of equation (\ref{Schroed})
generating a general solution $\Phi :\mathbb{R}\times \mathbb{R}\rightarrow
\mathbb{R}$\ of the Bohmian equation of motion (\ref{BMmotion}). Since
because of their uniqueness the maximal solutions do not intersect, we have $%
\Phi \left( t,\left( -\infty ,x\right) \right) =\left( -\infty ,\Phi \left(
t,x\right) \right) =\left( -\infty ,\gamma _{x}\left( t\right) \right) .$
From this it follows by means of equation (\ref{Tranport}) that%
\begin{equation}
\int_{-\infty }^{\gamma _{x}\left( t\right) }\rho \left( t,y\right)
dy=\int_{-\infty }^{x}\rho \left( 0,y\right) dy.  \label{Tranport_1}
\end{equation}%
As a check we may take the derivative of equation (\ref{Tranport_1}) with
respect to $t.$ This yields%
\[
\rho \left( t,\gamma _{x}\left( t\right) \right) \dot{\gamma}_{x}\left(
t\right) +\int_{-\infty }^{\gamma _{x}\left( t\right) }\partial _{t}\rho
\left( t,y\right) dy=0.
\]%
Making use of local probability conservation $\partial _{t}\rho =-\partial
_{x}j$ we recover by partial integration equation (\ref{BMmotion}).

Now observe that the equation (\ref{Tranport_1}) for $\gamma _{x}\left(
t\right) $ is meaningful not only when $\psi $ is a square integrable $C^{2}$%
-solution of equation (\ref{Schroed}) but also if $\psi \left( t,\cdot
\right) $ is an arbitrary representative of $\Psi _{t}$ with arbitrary $\Psi
_{0}\in L^{2}\left( \mathbb{R}\right) .$ In order to make this explicit let $%
E_{x}:L^{2}\left( \mathbb{R}\right) \rightarrow L^{2}\left( \mathbb{R}%
\right) $ with $x\in \mathbb{R}$ denote the spectral family of the position
operator. For the orthogonal projection $E_{x}$ holds
\[
\left( E_{x}\varphi \right) \left( y\right) =\left\{
\begin{array}{cc}
\varphi \left( y\right) & \text{for }y<x \\
0 & \text{otherwise}%
\end{array}%
\right. .
\]

The expectation value $\left\langle \Psi ,E_{x}\Psi \right\rangle $ of $%
E_{x} $ with a unit vector $\Psi \in L^{2}\left( \mathbb{R}\right) $ thus
yields the cumulative distribution function of the position probability
given by $\Psi .$ If we define $F:\mathbb{R}^{2}\rightarrow \left[ 0,1\right]
$ through $F\left( t,x\right) =\left\langle \Psi _{t},E_{x}\Psi
_{t}\right\rangle ,$ then equation (\ref{Tranport_1}) is equivalent to
\begin{equation}
F\left( t,\gamma _{x}\left( t\right) \right) =F\left( 0,x\right) .
\label{Tranport_2}
\end{equation}%
Thus, the graph $\left\{ \left( t,\gamma _{x}\left( t\right) \right)
\left\vert t\in \mathbb{R}\right. \right\} $ of a trajectory is a subset of
the level set of $F$ which contains the point $\left( 0,x\right) .$ If $\Psi
_{n}$ is a sequence in $L^{2}\left( \mathbb{R}^{s}\right) $ which converges
to $\Psi $ then
\begin{eqnarray*}
\lim_{n\rightarrow \infty }F_{n}\left( t,x\right) &=&\lim_{n\rightarrow
\infty }\left\langle \Psi _{n},e^{iht}E_{x}e^{-iht}\Psi _{n}\right\rangle
=\lim_{n\rightarrow \infty }\left\Vert E_{x}e^{-iht}\Psi _{n}\right\Vert ^{2}
\\
&=&\left\Vert E_{x}e^{-iht}\Psi \right\Vert ^{2}=F\left( t,x\right)
\end{eqnarray*}%
because $e^{-iht},E_{x},$ and $\left\Vert \cdot \right\Vert ^{2}$ are
continuous mappings.

Note that for any $t\in \mathbb{R}\ $the function $F\left( t,\cdot \right) :%
\mathbb{R}\rightarrow \left[ 0,1\right] $ is continuous and monotonically
increasing. Furthermore $\lim_{x\rightarrow -\infty }F\left( t,x\right) =0$
and $\lim_{x\rightarrow \infty }F\left( t,x\right) =1.$ The monotonicity is
a strict one if $\psi \left( t,\cdot \right) $ does not vanish on any
interval. Thus for any $\left( t,x\right) \in \mathbb{R}^{2}$ there exists
at least one $\gamma _{x}\left( t\right) \in \mathbb{R}$ such that equation (%
\ref{Tranport_2}) holds. (For those values $t$ for which $F\left( t,\cdot
\right) $ is strictly increasing, there exists exactly one $\gamma
_{x}\left( t\right) \in \mathbb{R}$ such that equation (\ref{Tranport_2})
holds.) The function $F$ cannot be constant in an open neighbourhood of some
point $\left( t,x\right) $ if the hamiltonian is bounded from below. Thus
for any $x\in \mathbb{R},$ for which there does not exist a neighbourhood on
which $F\left( 0,\cdot \right) $ is constant, we now \emph{define} $\gamma
_{x}:\mathbb{R}\rightarrow \mathbb{R}$ to be the unique \emph{continuous}
mapping for which
\[
F\left( t,\gamma _{x}\left( t\right) \right) =F\left( 0,x\right) .
\]%
Note that $\gamma _{x}:\mathbb{R}\rightarrow \mathbb{R}$ is continuous, yet
need not be differentiable.

\section{Hall's counter example}

Let us now illustrate this construction of not necessarily differentiable
Bohm\-ian trajectories by means of a solution $t\mapsto \Psi _{t}\notin
D_{h} $ of the Schr\"{o}dinger equation (\ref{SchroedQM}) describing a
particle confined to a finite interval on which the potential $V$ vanishes.
This solution has been used by Hall \cite{Hal} as a counter example to
Bohmian mechanics. Similar ones have been used in order to illustrate an
\textquotedblleft irregular\textquotedblright\ decay law $t\mapsto
\left\vert \left\langle \Psi _{0},\Psi _{t}\right\rangle \right\vert ^{2}.$
\cite{EF} Both works have made extensive use of Berry's earlier results
concerning this type of wave functions. \cite{Ber}

The (reduced) classical Schr\"{o}dinger equation corresponding to the
quantum dynamics is%
\begin{equation}
i\partial _{t}\psi \left( t,x\right) =-\partial _{x}^{2}\psi \left(
t,x\right)  \label{Schroed1}
\end{equation}%
for all $\left( t,x\right) \in \mathbb{R}\times \left[ 0,\pi \right] $
together with the homogeneous Dirichlet boundary condition $\psi \left(
t,0\right) =\psi \left( t,\pi \right) =0$ for all $t\in \mathbb{R}.$ The
corresponding hamiltonian's domain $D_{h}$ is the set of all those $\Psi \in
L^{2}\left( 0,\pi \right) $ which have an absolutely continuous
representative $\psi $ vanishing at $0$ and $\pi $ and whose distributional
derivatives up to second order belong to $L^{2}\left( 0,\pi \right) .$ As an
initial condition we choose the equivalence class of the function%
\[
\psi \left( 0,x\right) =1/\sqrt{\pi }\text{ for all }x\in \left[ 0,\pi %
\right] .
\]%
Since within the class $\Psi _{0}=\left[ \psi \left( 0,\cdot \right) \right]
$ there does not exist an absolutely continuous function vanishing at $0$
and $\pi $ the equivalence class $\Psi _{0}$ does not belong to $D_{h}.$ As
a consequence for any $t$ the vector $\Psi _{t}=e^{-iht}\Psi _{0}$ does not
belong to $D_{h}.$ This in turn implies that $\Psi _{t}$ does not have a
representative within the class of $C^{2}$-functions on $\left[ 0,\pi \right]
$ with vanishing boundary values.

The hamiltonian $h$ is self-adjoint. An orthonormal basis formed by
eigenvectors of $h$ is represented by the functions $u_{k}$ with
\[
u_{k}\left( x\right) =\sqrt{\frac{2}{\pi }}\sin \left( kx\right) \text{ for }%
0\leq x\leq \pi \text{ and }k\in \mathbb{N}.
\]

For $n\in \mathbb{N}$ the $C^{\infty }$-function $\psi _{n}:\mathbb{R}%
^{2}\rightarrow \mathbb{C}$ with%
\[
\psi _{n}\left( t,x\right) =\frac{4}{\pi \sqrt{\pi }}\sum_{k=0}^{n}\frac{%
e^{-i\left( 2k+1\right) ^{2}t}}{2k+1}\sin \left[ \left( 2k+1\right) x\right]
\]%
is a classical solution to the Schr\"{o}dinger equation (\ref{Schroed1}) on $%
\mathbb{R}^{2}$ and fulfills homogeneous Dirichlet boundary conditions at $%
x=0$ and $x=\pi .$ Furthermore $\psi _{n}$ is periodic not only in $x$ but
also in $t$ with period $2\pi .$ More precisely $\psi \left( t,\cdot \right)
$ is an odd trigonometric polynomial of degree $2n+1$ for any $t\in \mathbb{R%
}.$ In addition $\psi _{n}\left( t,\cdot \right) $ also is even with respect
to reflection at $\pi /2,$ i.e., it holds%
\[
\psi _{n}\left( t,\frac{\pi }{2}-x\right) =\psi _{n}\left( t,\frac{\pi }{2}%
+x\right)
\]%
for all $x\in \mathbb{R}.$ The functions $\psi _{n}\left( \cdot ,x\right) $
are trigonometric polynomials of degree $\left( 2n+1\right) ^{2}.$

\begin{figure}[h!]
\begin{center}
\includegraphics[scale=0.5]{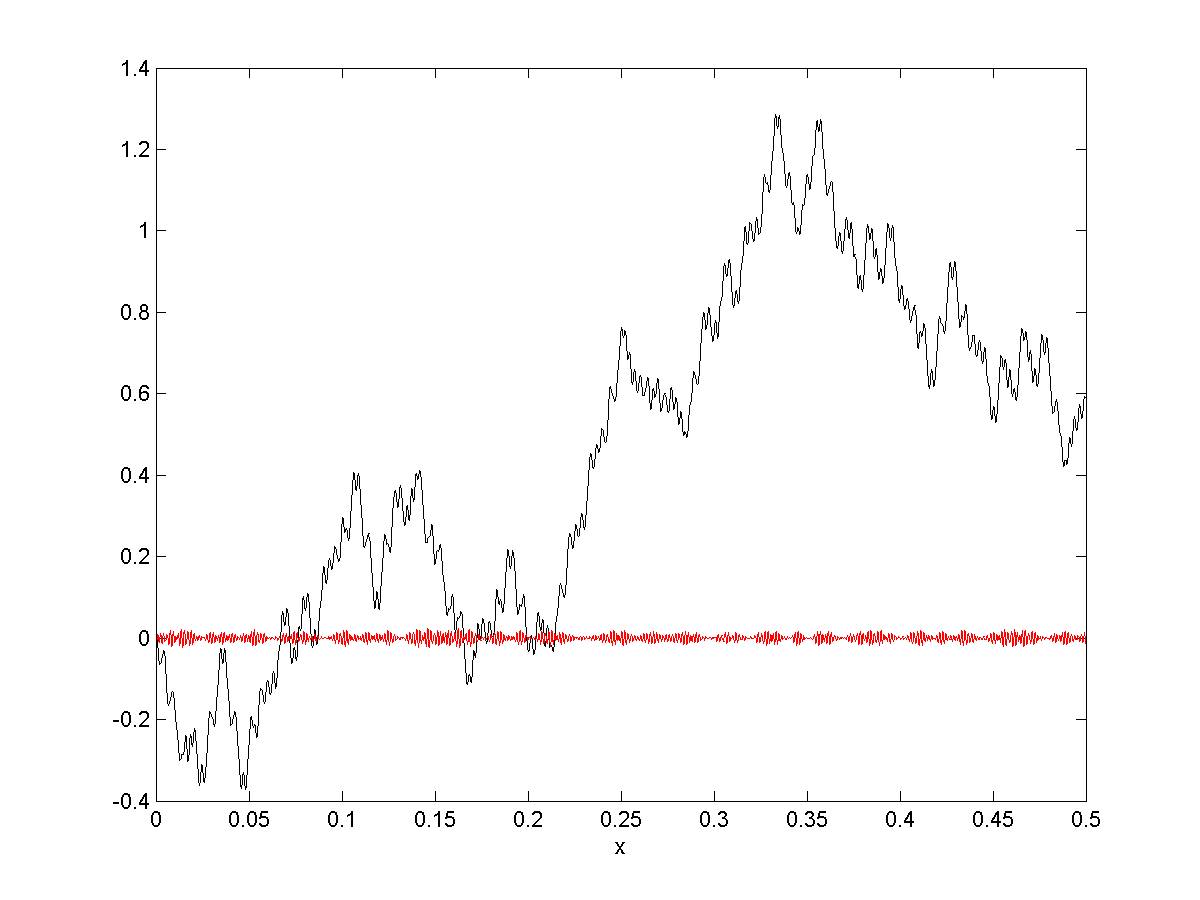}
\caption{Real part of $\protect\psi_{500}$ at a fixed time}
\label{psix}
\end{center}
\end{figure}

As is well known, the sequence $\left( \psi _{n}\left( 0,\cdot \right)
\right) _{n\in \mathbb{N}}$ converges pointwise on $\mathbb{R}.$ Its limit
is the odd, piecewise constant $2\pi $-periodic function $\sigma \left(
0,\cdot \right) $ with%
\[
\lim_{n\rightarrow \infty }\sqrt{\pi }\psi _{n}\left( 0,x\right) =\sqrt{\pi }%
\sigma \left( 0,x\right) =\left\{
\begin{array}{cc}
1 & \text{for }0<x<\pi \\
0 & \text{for }x\in \left\{ 0,\pi \right\}%
\end{array}%
\right. .
\]%
$\sigma \left( 0,\cdot \right) $ is discontinuous at $x\in \pi \cdot \mathbb{%
Z}.$ For any $t\in \mathbb{R}$ the sequence $\left( \psi _{n}\left( t,\cdot
\right) \right) _{n\in \mathbb{N}}$ converges pointwise on $\mathbb{R}$ to a
function $\psi \left( t,\cdot \right) .$ For rational $t/\pi $ this function
is piecewise constant. \cite{Ber} However for irrational $t/\pi $ the real-
and imaginary parts of $\psi \left( t,\cdot \right) $ restricted to any open
real interval have a graph with noninteger dimension. \cite{Ber} Thus for
irrational $t/\pi $ the function $\psi \left( t,\cdot \right) $ is
non-differentiable on any real interval. As an illustration we give in
Figure \ref{psix} the graph of
\[
x\mapsto \Re \sqrt{\pi }\psi _{500}\left( \frac{\pi }{\sqrt{12}},\pi
x\right)
\]%
for $0<x<1/2$ together with the partial sum over $k\in \left\{ 501,\ldots
750\right\} $ visible as the small noisy signal along the abscissa%
\[
x\mapsto \Re \sqrt{\pi }\left( \psi _{750}\left( \pi /\sqrt{12},\pi x\right)
-\psi _{500}\left( \pi /\sqrt{12},\pi x\right) \right) .
\]

Similarly, for given $x\in \left( 0,\pi \right) $ the mapping
$t\mapsto \psi
\left( t,x\right) $ does not belong to the set of piecewise $C^{1}$%
-functions on $\left[ 0,2\pi \right] .$ This can be seen as follows. First
note that for given $x$ the $2\pi $-periodic function $\psi \left( \cdot
,x\right) $ has the Fourier expansion%
\begin{eqnarray*}
\psi \left( t,x\right) &=&\sum_{k=1}^{\infty }c_{n}e^{-int}\text{ where} \\
c_{n} &=&\left\{
\begin{array}{ll}
\frac{4}{\pi \sqrt{\pi }}\frac{1}{2k+1}\sin \left[ \left( 2k+1\right) x%
\right] & \text{for }n=\left( 2k+1\right) ^{2}\text{ with }k\in \mathbb{N}
\\
0 & \text{otherwise}%
\end{array}%
\right. .
\end{eqnarray*}%
Assume now that $\psi \left( \cdot ,x\right) $ is piecewise $C^{1}.$ Then,
according to a well known property of Fourier coefficients, there exists a
positive real constant $C$ such that $n\left\vert c_{n}\right\vert <C$ for
all $n\in \mathbb{N}.$ This implies that%
\begin{equation}
\left( 2k+1\right) \left\vert \sin \left[ \left( 2k+1\right) x\right]
\right\vert <C^{\prime }\text{ for all }k\in \mathbb{N}
\label{decayestimate}
\end{equation}%
with the positive constant $C^{\prime }=\pi \sqrt{\pi }C/4.$ However, for $%
x\notin \pi \cdot \mathbb{Z}$ there exists some real constant $\varepsilon
>0 $ such that the set $\left\{ k\in \mathbb{N}:\left\vert \sin \left[
\left( 2k+1\right) x\right] \right\vert >\varepsilon \right\} $ contains
infinitely many elements. Thus for $x\notin \pi \cdot \mathbb{Z}$ the
estimate (\ref{decayestimate}) cannot hold and therefore the function $%
t\mapsto \psi \left( t,x\right) $ cannot be piecewise $C^{1}$ on $\left[
0,\pi \right] .$

\begin{figure}[h!]
\begin{center}
\includegraphics[scale=0.5]{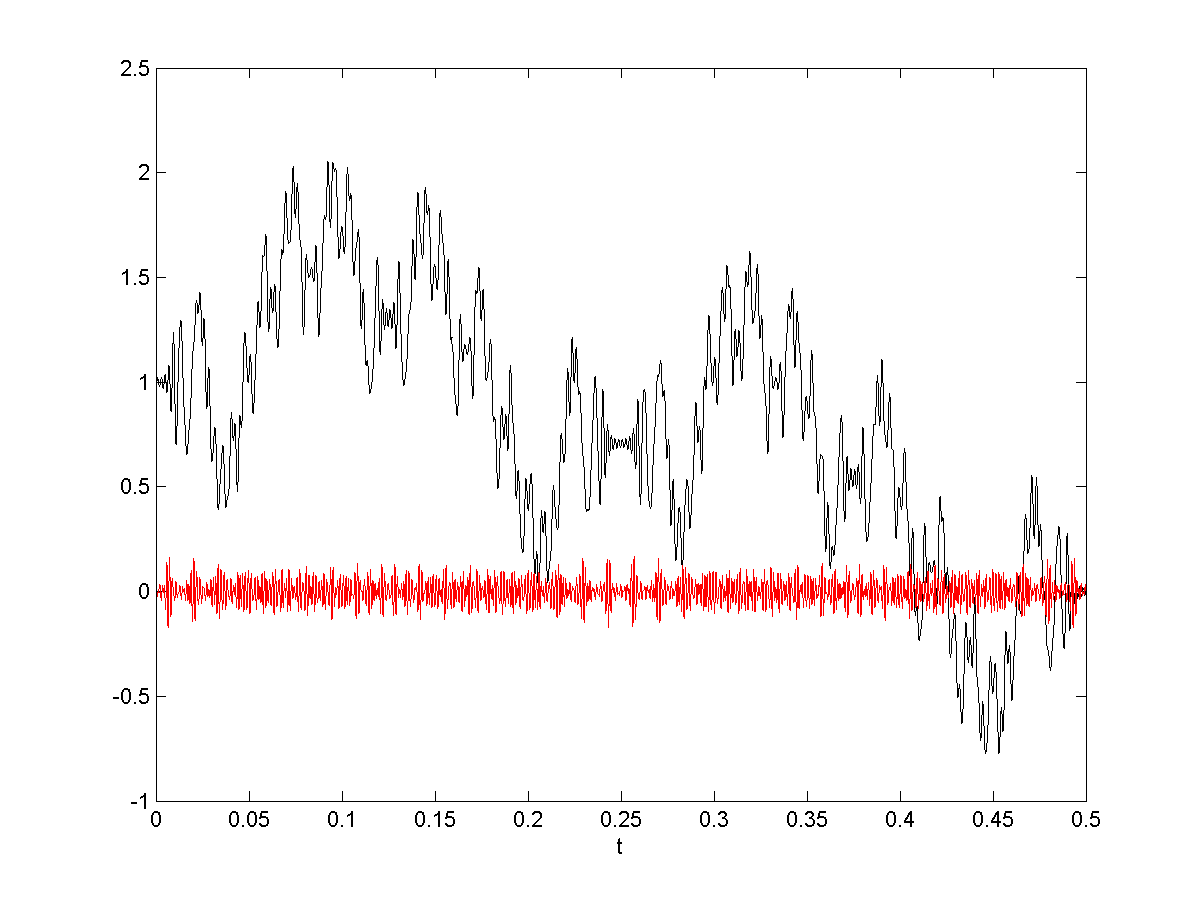}
\caption{Real part of $\protect\psi_{15}$ at $x=\protect\pi /2$}
\label{PsiT}
\end{center}
\end{figure}

In Figure \ref{PsiT} we plot the time dependence%
\[
t\mapsto \Re \sqrt{\pi }\psi _{15}\left( \pi t,\pi /2\right) =\frac{4}{\pi }%
\sum_{k=0}^{15}\frac{\left( -1\right) ^{k}}{2k+1}\cos \left( \left(
2k+1\right) ^{2}\pi t\right)
\]%
for $0<t<1/2$ together with the partial sum
\[
t\mapsto \Re \sqrt{\pi }\left( \psi _{20}\left( \pi t,\pi /2\right) -\psi
_{15}\left( \pi t,\pi /2\right) \right)
\]
(noisy signal along abscissa).

The restriction of the limit $\sigma \left( 0,\cdot \right) $ to
$\left[ 0,\pi \right] $ represents the same $L^{2}$ element as $\psi
\left( 0,\cdot \right) $ does. Thus $\lim_{n\rightarrow \infty
}\left\Vert \left[ \widetilde{\psi _{n}}\left( 0,\cdot \right)
\right] -\Psi _{0}\right\Vert =0, $ when $\widetilde{\psi _{n}}$
denotes the restriction of $\psi _{n}$ to $\mathbb{R}\times \left[
0,\pi \right] .$ Correspondingly $\rho _{n}\left( 0,\cdot \right)
=\left\vert \psi _{n}\left( 0,\cdot \right) \right\vert ^{2}$
converges towards the density of the equipartition on $\left[ 0,\pi
\right] . $ Additionally, due to the continuity of the evolution
operator $e^{-iht},$ the sequence of equivalence classes $\left[
\widetilde{\psi _{n}}\left( t,\cdot \right) \right] \in
C_{h}^{\infty }$ approximates the $L^{2}$ vector
$\Psi _{t}=e^{-iht}\Psi _{0},$ i.e., for all $t$ holds%
\[
\lim_{n\rightarrow \infty }\left\Vert \left[ \widetilde{\psi _{n}}\left(
t,\cdot \right) \right] -\Psi _{t}\right\Vert =0.
\]

Since also $E_{x}:L^{2}\left( \mathbb{R}\right) \rightarrow L^{2}\left(
\mathbb{R}\right) $ is continuous, the time dependent cumulative position
distribution function $F:\mathbb{R}\times \left[ 0,\pi \right] \rightarrow %
\left[ 0,1\right] $ with $F\left( t,x\right) :=\left\langle \Psi
_{t},E_{x}\Psi _{t}\right\rangle $ obeys%
\[
F\left( t,x\right) =\lim_{n\rightarrow \infty }\left\langle \left[
\widetilde{\psi _{n}}\left( t,\cdot \right) \right] ,E_{x}\left[ \widetilde{%
\psi _{n}}\left( t,\cdot \right) \right] \right\rangle =\lim_{n\rightarrow
\infty }\int_{0}^{x}\left\vert \psi _{n}\left( t,y\right) \right\vert
^{2}dy.
\]

The level lines of the functions $F_{n}:\mathbb{R}\times \left[ 0,\pi \right]
\rightarrow \left[ 0,1\right] $ with $F_{n}\left( t,x\right)
:=\int_{0}^{x}\left\vert \psi _{n}\left( t,y\right) \right\vert ^{2}dy$ thus
converge to the continuous level lines of $F.$

\begin{figure}[h!]
\begin{center}
\includegraphics[scale=0.5]{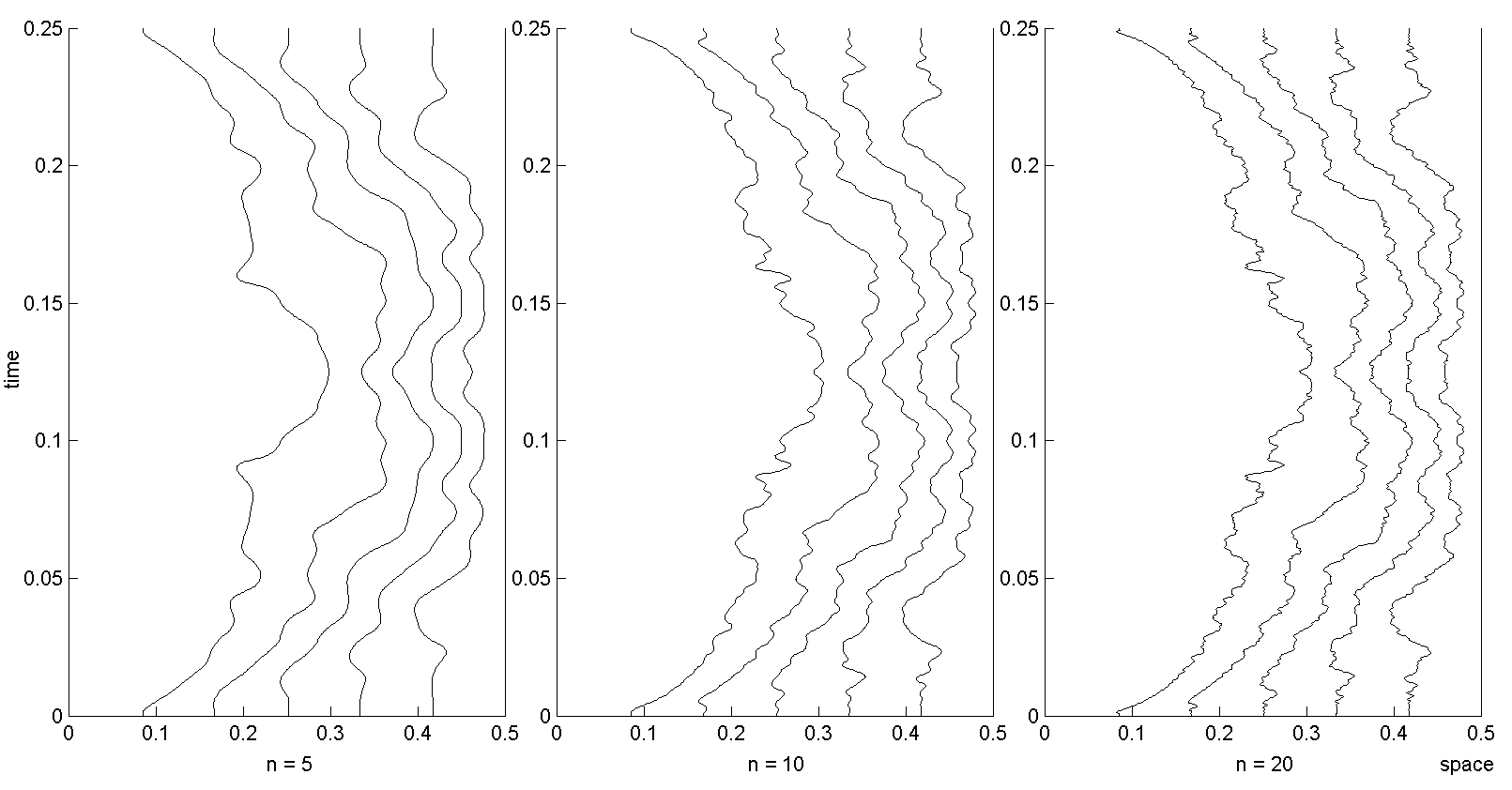}
\caption{Level lines of $F_{n}$ for $n=5,10,20$}
\label{traj_app}
\end{center}
\end{figure}

Figure \ref{traj_app} shows some level lines of $F_{n}$ for $n=5,10,20$
starting off at equal positions at $t=0.$ The level lines inherit the period
$\pi /4$ of $F\left( \cdot ,x\right) ,$ which has this periode since the
frequencies appearing in the even function $\left\vert \psi _{n}\left( \cdot
,x\right) \right\vert ^{2}$ are $0,8,16,\ldots $

\begin{figure}[h!]
\begin{center}
\includegraphics[scale=0.6]{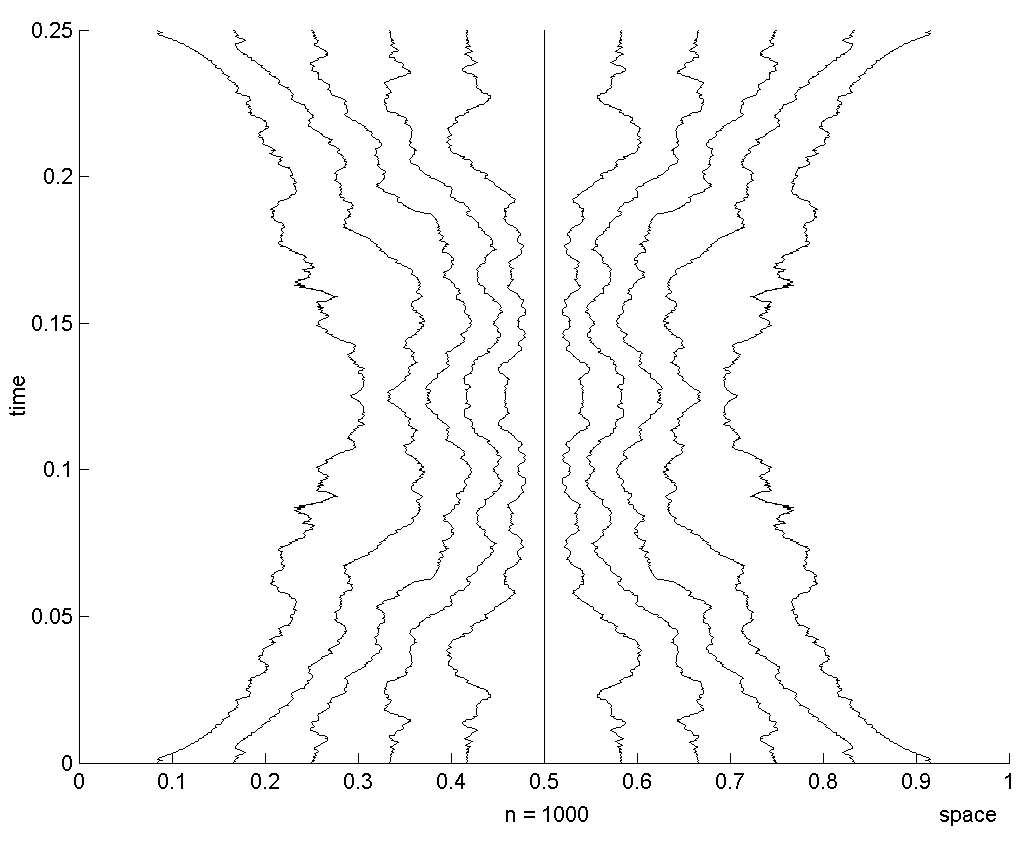}
\caption{Level lines of $F_{1000}$} \label{Traj1000}
\end{center}
\end{figure}

Figure \ref{Traj1000} shows the case $n=1000.$ Increasing $n$ from $20$ to $%
1000$ hardly alters the level lines.

\end{document}